\documentstyle[12pt]{article}
\catcode`\@=11
\def\maketitle{\par
 \begingroup
 \def\thefootnote{\fnsymbol{footnote}}
 \def\@makefnmark{\hbox
 to 0pt{$^{\@thefnmark}$\hss}}
 \if@twocolumn
 \twocolumn[\@maketitle]
 \else \newpage
 \global\@topnum\z@ \@maketitle \fi\thispagestyle{empty}\@thanks
 \endgroup
 \setcounter{footnote}{0}
 \let\maketitle\relax
 \let\@maketitle\relax
 \gdef\@thanks{}\gdef\@author{}\gdef\@title{}\let\thanks\relax}
\def\@maketitle{\newpage
 \null
 \hbox to\textwidth{\hfil\hbox{\begin{tabular}{r}\@preprint\end{tabular}}}
 \vskip 2em \begin{center}
 {\Large\bf \@title \par} \vskip 1.5em {\normalsize \lineskip .5em
\begin{tabular}[t]{c}\@author
 \end{tabular}\par}
 \end{center}
 \par
 \vskip 1.5em}
\def\preprint#1{\gdef\@preprint{#1}}
\catcode`\@=12

\begin{document}
\baselineskip=.32in

\preprint{DPNU-94-42\\ SNUTP-94-98\\ hepth@xxx/9410119}
\title{\large\bf Comment on ``Obstructions to the Existence of Static Cosmic
Strings in an Abelian Higgs Model"\protect\\[1mm]\  }
\author{}
\author{\normalsize Chanju Kim$^{\ast}$\\
{\normalsize\it Center for Theoretical Physics and
Department of Physics}\\
{\normalsize\it Seoul National University,
Seoul 151-742, Korea}\\[4mm]
\normalsize Yoonbai Kim$^{\dagger}$\\
{\normalsize\it Department of Physics, Nagoya University}\\
{\normalsize\it Nagoya 464-01, Japan}}
\date{}
\maketitle

\newpage

\indent There has been considerable interest in the possible cosmic string
configurations and Y.\ Yang has recently derived intriguing obstructions to
the existence of cylindrically symmetric solutions in the Bogomol'nyi
critical phase of the Einstein-Maxwell-Higgs system \cite{Yan}.
In this comment, we would like to mention some earlier results
and recent progress \cite{Lin,Kim} which were not reflected in Ref.\
\cite{Yan} but are important in connection with the obstructions he
claimed. To our opinion,
some of his obstructions are not consistent with the results of Refs.\
\cite{Lin,Kim} which we briefly summarize below. Also, we give a new
condition stronger than those found before.

In the self-dual Abelian Higgs model coupled with gravity, once we
restrict our interest to
the string configurations parallel to the $z$ axis and satisfying  Bogomol'nyi
equations, Einstein equations turn out to be solved under the general
stationary metric and the whole solutions can be described by a single
second-order equation for the amplitude of the Higgs field \cite{CG}
\begin{eqnarray} \label{1}
\lefteqn{\partial^{2}\ln\frac{f^{2}}{\prod_{p=1}^{n}|z-z_{p}|^{2}}}\nonumber\\
 &&=e^{h+\bar{h}}
\Bigl(\frac{f^{2}e^{-(f^{2}-1)}}{\prod_{p=1}^{n}|z-z_{p}|^{2}}
\Bigr)^{4\pi Gv^{2}}(f^{2}-1),
\end{eqnarray}
where $f$ is the dimensionless scalar amplitude rescaled by the vacuum
expectation
value $v$, $h(z)$ ($\bar{h}(\bar{z})$) is a holomorphic (antiholomorphic)
function which reduces to a constant, say $F$, in the cylindrically symmetric
case, and $z_{p}$ represents the center of $p$-th vortex .
[The gravitational
constant $G$ in our notation corresponds to $G/2$ of Ref.\ \cite{Yan}.]

The finite energy condition for $f$ is not simply that of flat case,
$f(\infty)=1$, but more complicated. For cylindrically symmetric case, it is
given by
\begin{equation} \label{2}
f(\infty)=\left\{\begin{array}{l}
\mbox{1 or 0 if $0<4\pi Gv^{2}n\le1$}\\
f_\infty,\ 0\le f_\infty\le1\ \mbox{if }4\pi Gv^{2}n>1,
\end{array}
\right.
\end{equation}
where $n$ is the vorticity of the vortices superimposed at the origin (i.e.,
$f\sim f_0r^n$ near the origin). Other than the well-known conic solutions
which corresponds to $f(\infty)=1$ with $0<4\pi Gv^{2}n<1$,
non-conic solutions were first found by Linet \cite{Lin} for $4\pi Gv^{2}n=1$.
In this case Eq.\ (\ref{1}) (for the cylindrically symmetric case) is
integrated to the first-order equation
\begin{equation}
\frac{df}{dr}=\frac{f}{r}\sqrt{n^{2}-\frac{Fn}{2}f^{\frac{2}{n}}
e^{-(f^{2}-1)}}.
\end{equation}
Based on this, one can obtain the solutions for which
the underlying two-manifold $\Sigma$ is an (noncompact) asymptotic cylinder
if $f(\infty)=1$ with $F=2n$ and $\Sigma\approx S^2$ (compact)
if $f(\infty)=0$ with $n=1$ and $F>2$. Here, the condition $n=1$ which was
overlooked in Refs.\ \cite{Lin,Kim} is necessary for the regularity of the
solutions on the whole manifold $\Sigma\approx S^2$ and hence there exist
only $N=2$ solutions on $S^2$. [One may easily convince himself that,
otherwise, the phase of the Higgs field would not be regular everywhere
outside the vortex points.] The maximum circumference of the
cylinder and that of two-sphere along the tropical line has a fixed value
$\sqrt{32\pi^3G}n/e$ independently of $F$
\cite{Kim}. There is no other type of regular solutions to Eq.\
(\ref{1}) in a geodesically complete manifold \cite{Kim}.
[In fact, in  Ref.\ \cite{Kim}, we considered more general system which
may have external point particles.]
That is, among the many possibilities in Eq.\ (\ref{2}),
regular finite-energy solutions exist only for the aforementioned three
cases.\newline
\indent From these, we see that
(i) in Ref.\ \cite{Yan}, the second condition of Eq.\ (7) in Theorem 2 and
Eq.\ (14) in Theorem 5 (and subsequent discussions) can not be true since
there exist asymptotic-cylinder solutions;
(ii) contrary to Yang's claim of the existence of $|N|\ge 3$ solutions for
$\Sigma\approx S^2$, regular solutions exist only for $N=2$.
Correspondingly, in this case, vacuum expectation value $v$ is quantized
uniquely as $v=1/\sqrt{4\pi G}$ rather than $v$ can take arbitrary value of
Eq.\ (4) of Ref.\ \cite{Yan}.

\vspace{5mm}

We thank Choonkyu Lee for reading the manuscript.
This work was supported by the Korea Science and Engineering Foundation
through Center for Theoretical Physics, Seoul National University
and JSPS under $\#$93033 (Y.K.).


\newpage

\def\hebibliography#1{\begin{center}\subsection*{References
}\end{center}\list
  {[\arabic{enumi}]}{\settowidth\labelwidth{[#1]}
\leftmargin\labelwidth	  \advance\leftmargin\labelsep
    \usecounter{enumi}}
    \def\newblock{\hskip .11em plus .33em minus .07em}
    \sloppy\clubpenalty4000\widowpenalty4000
    \sfcode`\.=1000\relax}

\let\endhebibliography=\endlist

\begin{hebibliography}{10}
\item[$\ast$] E-mail address: cjkim$@$phyb.snu.ac.kr
\item[$\dagger$] E-mail address: yoonbai$@$eken.phys.nagoya-u.ac.jp
\bibitem{Yan} Y.\ Yang, Phys.\ Rev.\ Lett.\ {\bf 73}, 10 (1994).
\bibitem{Lin} B.\ Linet, Class.\ Quantum Grav.\ {\bf 7} (1990) L79.
\bibitem{Kim} C.\ Kim,\ and Y.\ Kim, Phys.\ Rev.\ {\bf 50}, 1040 (1994).
\bibitem{CG} B.\ Linet, Gen.\ Rel.\ Grav.\ {\bf 20} (1988) 451; A.\ Comtet,
and G.\ W.\ Gibbons, Nucl.\ Phys.\ {B299} (1988) 719.
\end{hebibliography}
\end{document}